\documentclass[useAMS,usenatbib]{mn2e}
\usepackage{times}
\usepackage{latexsym}
\usepackage{graphicx}

\title[WMAP constraints on low redshift evolution of dark energy]{WMAP constraints on low redshift evolution of dark energy}

\author[Jassal, Bagla and Padmanabhan]{H. K. Jassal
  $^{1,2}$ \thanks{E-mail: hkj@mri.ernet.in}, J.~S.~Bagla $^2$
  \thanks{E-mail: jasjeet@mri.ernet.in} and 
  T.~Padmanabhan$^1$ \thanks{E-mail: nabhan@iucaa.ernet.in} \\
$^{1}$ Inter University Centre for Astronomy and Astrophysics,
Post Bag 4, Ganeshkhind, Pune 411 007, India. \\
$^{2}$ Harish-Chandra Research Institute, Chhatnag Road,
Jhunsi, Allahabad 211 019, India.}

\begin{document}
\date{} 

\pagerange{\pageref{firstpage}--\pageref{lastpage}} \pubyear{2004}
\maketitle
\label{firstpage}
\begin{abstract}
The conceptual difficulties associated with a cosmological constant have led
to the investigation of alternative models in which the equation of state
parameter, $w=p/\rho$, of the dark energy evolves with time. 
We show that combining the  supernova type Ia observations {\it with the
  constraints from WMAP observations} restricts large variation of
$\rho(z)$ at low redshifts. 
The combination of these two observational constraints is stronger than either
one. 
The results are completely consistent with the cosmological constant as the
source of dark energy. 
\end{abstract}

\begin{keywords}
Cosmology: cosmic microwave background, cosmological parameters.
\end{keywords}

\section{Introduction}

Well before the data from the high redshift supernova project became
available, several independent constraints indicated the existence of a
cosmological constant \citep{crisis1,crisis2,crisis3}. 
In the last decade, observational evidence for an accelerating universe has
become conclusive with almost all other possibilities being ruled out by
observations of high redshift supernovae
\citep{nova_data1,nova_data2,nova_data3} and the cosmic 
microwave background radiation (CMBR) \citep{boomerang,wmap_params}. 
The accelerated expansion of the universe requires either a cosmological
constant or some form of dark energy
\citep{review1,review2,review3,review4} to drive the acceleration, with
$w\equiv p/\rho < -1/3$.  
Although a cosmological constant is the simplest solution from a
phenomenological point of view (requiring just one fine tuned parameter),
there is no natural explanation of the small observed value. 
This has led theorists to develop models in which a field,
typically a scalar field, provides the source of dark energy, e.g.,
quintessence
\citep{quint1,quint2,quint3,quint4,quint5,quint6,quint7,quint8},
k-essence
\citep{k-essence1,k-essence2,k-essence3,k-essence4,k-essence5},
tachyons
\citep{tachyon1,tachyon2,tachyon3,tachyon4,tachyon5,tachyon6,tachyon7,tachyon8,tachyon9,tachyon10,tachyon11,tachyon12},
phantom fields
\citep{phantom1,phantom2,phantom3,phantom4,phantom5,phantom6,phantom7,phantom8,phantom9,phantom10,phantom11},
branes
\citep{brane1,brane2,brane3,brane4,brane5},  
etc.  
In the absence of significant {\it spatial} variation in the dark energy, the
key difference between such models and the one with the cosmological constant
is that, in general, $w$ is a function of redshift $z$ in the former.  
Perturbations in dark energy also lead to observable signatures, though these
can easily be confused with other physical effects
\citep{perturb1,perturb2,perturb3}. 
There have also been a few proposals for unified dark matter and dark
energy \citep{unified_dedm1,unified_dedm2,unified_dedm3} but these
models are yet to be  developed in sufficient detail to allow direct
comparison with observations in a fruitful manner.  
If the current observations had excluded $w=-1$ then one could have
immediately ruled out cosmological constant as a candidate; but since this is
not the case, direct exploration of $w(z)$ at different redshifts, in order to 
check possible dependence of $w$ on  $z$, is of importance.  
Observations constrain the entire suite of parameters that describe
cosmological parameters and while it is possible to choose other cosmological
parameters so that $\Lambda$CDM is not allowed, these models have not been
ruled out so far.

It has been known for some time that supernova observations and constraints
from structure formation can be combined to put stringent limits on models for
dark energy \citep{lss_de}.  
Several attempts have been made in the past to constrain the equation of state 
for dark energy, along with  other cosmological parameters, using the
observations of galaxy clustering, temperature anisotropies in the CMBR and
the high redshift supernovae
\citep{lss_de,constraints_1,constraints_2,constraints_3,constraints_4,constraints_5,constraints_7,constraints_8,constraints_9,constraints_10}. 

This work, which is in the same spirit, focuses exclusively on constraining
the variation of equation of state for dark energy by using a combination of
such observations.  
We wish to constrain the variation of dark energy while keeping most of the
other cosmological parameters fixed around their favoured values. 
In particular, we study the effect of a varying equation of state on
the angular power spectrum of the  CMBR fluctuations.  
Dark energy is not expected to be dynamically significant  at the time of
decoupling and --- in fact --- models in which this is not true are plagued by 
slow growth of density perturbations
\citep{quint_growth1,quint_growth2,quint_growth3}. 
Nevertheless, evolving dark energy will affect the features of temperature
anisotropies in the CMBR in at least two ways: (1)~The angular scale of 
features in temperature anisotropy, like the peaks, will change since
the angular diameter distance depends on the form of $w(z)$. 
(2)~The integrated Sachs-Wolfe (ISW) effect will also depend on the nature of
dark energy and its evolution, this effect is more relevant at small $l$.
Thus observations of temperature anisotropies in the CMBR can be used to
constrain the {\it evolution} of dark energy. 
Combined with the supernova observations, this allows us to put tight
constraints on the equation of state of dark energy and its evolution. 
Our approach here is to use the full WMAP angular power spectrum in order to
ensure that both the effects mentioned above are captured in the analysis.  
For lower multipoles ($l \leq 20$) signals from the ISW effect, perturbations
in dark energy and reionisation need to be disentangled, however the
relative importance of this part of the angular power spectrum is limited as
we use the full angular power spectrum from WMAP.  
Our aim is to demonstrate that the combination of WMAP observations and high
redshift supernova observations is a very powerful constraint on variations in
dark energy, certainly more powerful that either of the observations used in
isolation.  
{\it As far as we know, most previous attempts to constrain the dark energy
  sector using WMAP data have not used the full angular power spectrum. }

\section{Varying Dark Energy}

Supernova observations, as well as the angular power spectrum of temperature
anisotropies in the microwave background put geometric constraints on dark
energy.  
Thus these observations constrain the Hubble parameter $H(z)$, which for the
models under consideration here can be written as:
\begin{equation}
H^2(z) = H_0^2 \left[ \Omega_{nr} \left(1+z\right)^3 +
  \varrho^{DE}(z)/\varrho^{DE}_0 \right] \nonumber 
\end{equation}
where we have ignored the contribution of radiation to energy density.
Observed fluxes from supernovae and location of peaks in the CMB constrain
distances to supernovae and to the surface of last scattering \citep{white},
respectively.  
Dark energy can be constrained only at redshifts where it contributes
significantly to the energy density, therefore observations are sensitive to
changes in $\varrho^{DE}(z)$ at low redshifts and  variations at high
redshifts ($z \gg 1$) will be difficult to detect.  
Variation in $\varrho^{DE}(z)$ can be written in terms of changes in the
equation of state parameter $w$.  
Once we allow for variation of $w$, one is dealing with a {\it function}
which, technically, has an infinite number of parameters.  
Given a finite number of observations, one can always fine tune such a
function.  
From a practical point of view, it is necessary (and often sufficient) to
represent $w(z)$ using a small number of parameters and do the analysis. 
The results will necessarily have some amount of parameterization dependence
\citep{dynamic_de14} but this can be controlled by choosing different forms of
parameterization and determining the range of variations in the results. 
In this work, we shall use the two parameterizations:
\begin{equation}
w(z) = w_0 +  w_1 \frac{z}{(1+z)^p}; \,\,\,\; p=1,2 \label{taylor}
\end{equation}
For both $p=1,2$ we have $w(0)=w_0;w'(0)=w_1$ but the high redshift behaviour
of these functions are different: $w(\infty)=w_0+w_1$ for $p=1$ while
$w(\infty)=w_0$ for $p=2$. 
Hence $p=2$ can model a dark energy component which has the same equation of
state at the present epoch and at high redshifts, with rapid variation at low
$z$. 
Observations are not very sensitive to variations in w(z) for $z \gg 1$, hence
return to the present value is not of critical importance.  
However, it does allow us to probe rapid variations at small redshifts.  
For $p=1$, we can trust the results only if $w_0 + w_1$ is well below zero at
the time of decoupling so that dark energy is not relevant for the physics of
recombination of the evolution of perturbations up to that epoch.  
As we shall see, the allowed range of parameters does not contain such
models. 
This constraint on the asymptotic value means that this parameterisation
cannot be used to probe rapid variations at small redshifts.

We will take $\Omega_{de} + \Omega_{nr} =1$ \citep{boomerang,wmap_params} where
density parameter for the total non-relativistic component is denoted by
$\Omega_{nr}$ and the density parameter for dark energy is $\Omega_{de}$.  
We will make assumptions regarding a few other parameters as well:
(i) We assume that $\Omega_B = 0.05$ and (ii) $h=0.7$. 
(iii) We also assume that the density fluctuations have the
Harrison-Zel'dovich spectrum, i.e., $n=1$. 
We do this since (a) there have been numerous studies in the past, restricting
the standard cosmological parameters and (b) we wish to focus on the effect of
$w_1\neq 0$. 
One can quarrel with these assumptions but these are sufficient to give us an
idea of the parameter space available for variation of dark energy. 
A complete study of these issues will need to address two questions: what is
the parameter space available for variation of dark energy if all parameters
of interest in cosmological models are allowed to vary, this essentially
addresses the issue of potential degeneracies of variation of dark energy with
other parameters, and, if dark energy is allowed to vary then does that
increase or change the allowed range of other parameters?  
We are carrying out a detailed study to address these issues \citep{jbp2}.  
We will comment on dependence of our results on some of these parameters
whenever relevant. 

For the supernova analysis, which is by now standard
\citep{dynamic_de1,dynamic_de2,dynamic_de3,dynamic_de4,dynamic_de5,dynamic_de6,dynamic_de7,dynamic_de8,dynamic_de9,dynamic_de10,dynamic_de11,dynamic_de12,dynamic_de13,dynamic_de14,dynamic_de15},
we combine the data on $230$ supernovae published by Tonry et al. and also for
the $23$ supernovae listed by Barris et
al. \citep{nova_data1,nova_data2,nova_data3}.    
To account for the uncertainty in measurement of cosmological redshift, we
neglect the data points at redshift $z<0.01$.   
The points with $A_V > 0.5$ are neglected because of the uncertainty in host
galaxy extinction, after these cuts we are left with data for $194$
supernovae. 
We compare the distance modulus predicted by theoretical models to
the observed variation of apparent magnitudes of high redshift supernovae.   
We use the $\chi^2$ minimization method to find the best fit model.  
The supernova data favours dark energy to have an equation of state $-1.3 \leq
w(0) \leq -0.7$ at the present epoch, with the lower values favoured for
larger $\Omega_{nr}$.  
Significant variation in energy density of dark energy is allowed by supernova
observations (see Figure~\ref{fig:wenvelope}), e.g. the energy density can
change by more than a factor of $3$ by $z=1$.
The results we obtain are consistent with similar analysis in other papers
 \citep{dynamic_de1,dynamic_de2,dynamic_de3,dynamic_de4,dynamic_de5,dynamic_de6,dynamic_de7,dynamic_de8,dynamic_de9,dynamic_de10,dynamic_de11,dynamic_de12,dynamic_de13,dynamic_de14,dynamic_de15}.   
Use of supernovae sorted on the basis of data quality
\citep{nova_data1,nova_data2,nova_data3} gives similar, though somewhat
tighter constraints.   

Let us now add constraints from the WMAP observations.
The location of peaks in the angular power spectrum of temperature
anisotropies (which depends on the angular diameter distance, which --- in
turn --- depends on $w(z)$) provides the main constraint.
Other effects like ISW provide a further constraint but their impact is
limited as these are restricted to $l \leq 20$ for most models whereas we use
the full WMAP angular power spectrum ($l \leq 900$). 
Given that the physics of recombination and evolution of perturbations
does not change if we only modify $w(z)$ within some safe limits, any
change in the observed angular power spectrum will be due to a change
in $w(z)$.
The location of peaks in the angular power spectrum of temperature
anisotropies will provide the main constraint and hence fix the effective
value of $w(z)$.
We can define a $w_{eff}$ by requiring that the angular size of a critical
scale like the Hubble radius at the time of decoupling is same in the model
with a given $w(z)$ and for the model where $w(z)=w_{eff}=$~constant. 
Observations mainly constrain the value of $w_{eff}$. 
Thus if the present value $w_0 < w_{eff}$ then we should have $w_1 > 0$, and
similarly if $w_0 > w_{eff}$ then $w_1 < 0$ in the parameterisations we have
used. 
Of course, if $w_1=0$ then $w_{eff}=w_0$.

The priors we have considered here is $H_0=70$km~s$^{-1}$~Mpc$^{-1}$, $n=1$,
$\Omega_B=0.05$, $\tau=0$, $-1.5 \le w_0 \le -0.7$,  $-2 \le w_1 \le 2$ and
$0.2 \le \Omega_M \le 0.4$.
We do not vary many of these parameters as our main aim is to study the
allowed variation in dark energy.
If we allow variation of all parameters at once, it will become difficult to
understand the impact of individual variations. 
(We have varied some of the parameters individually, e.g., we have checked that
allowing for reionisation with a prior $\tau \leq 0.4$ does not make a
significant difference to the results, even though the best fit value of
optical depth $\tau$ is different for different models.) 
We are carrying out a detailed study where variation of most of these
parameters is allowed and we also bring in more constraints \cite{jbp2}. 

We do not take into account perturbations in dark energy in this paper. 
Perturbations in dark energy can modify our results to some extent.
The effect of perturbtations depends on the detailed model of dark energy and
cannot be incorporated in the model independent aprroach we have used here. 
Including adiabatic perturbations in dark energy does not change our result
significantly as the effect of these perturbations changes the angular power
spectrum only for $l \leq 20$. 
The effect of perturbations in dark energy becomes more relevant if we
consider the allowed values of all cosmological parameters in a varying dark
energy model.  

To bring in these constraints from WMAP data, we need to compare the
angular power spectrum expected in a given model with the observed
spectrum \citep{wmap_cl}. 
We compute the theoretical angular power spectrum for each model using
CMBFAST \citep{cmbfast}.
The comparison is carried out using the likelihood program made
available by the WMAP team  \citep{wmap_lik}.
We normalize the angular power spectrum by maximizing likelihood in a
comparison with the WMAP data. 

\begin{figure}
\begin{center}
\includegraphics[width=8.0cm]{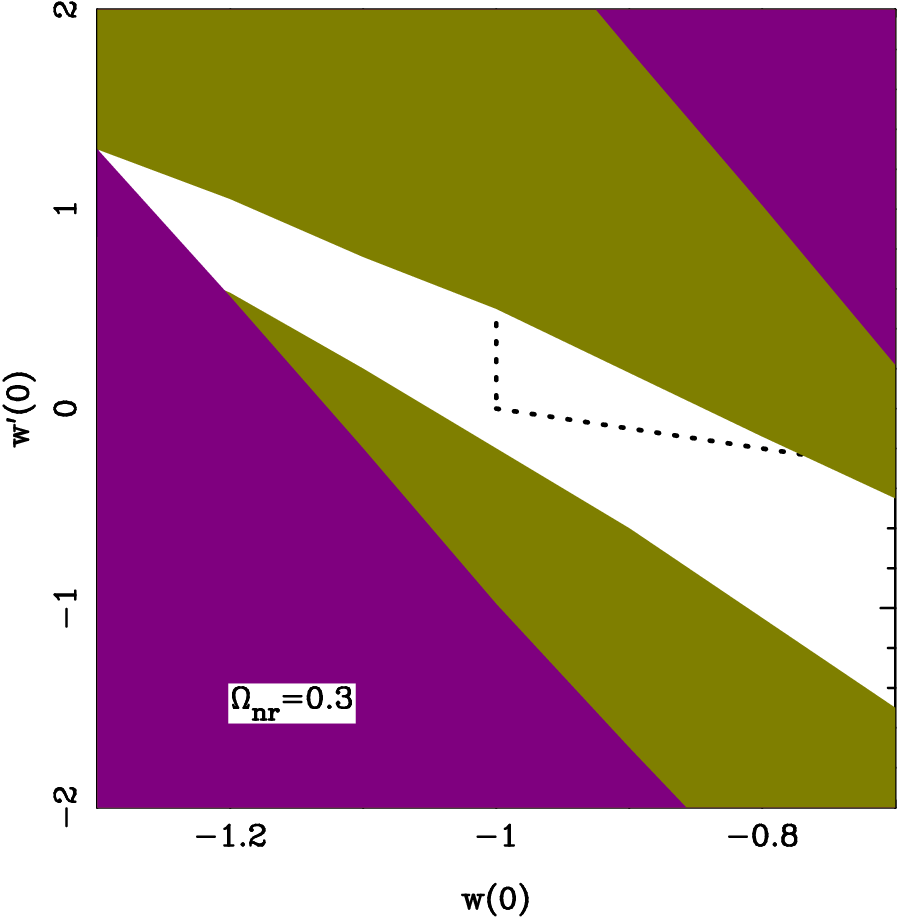}
\includegraphics[width=8.0cm]{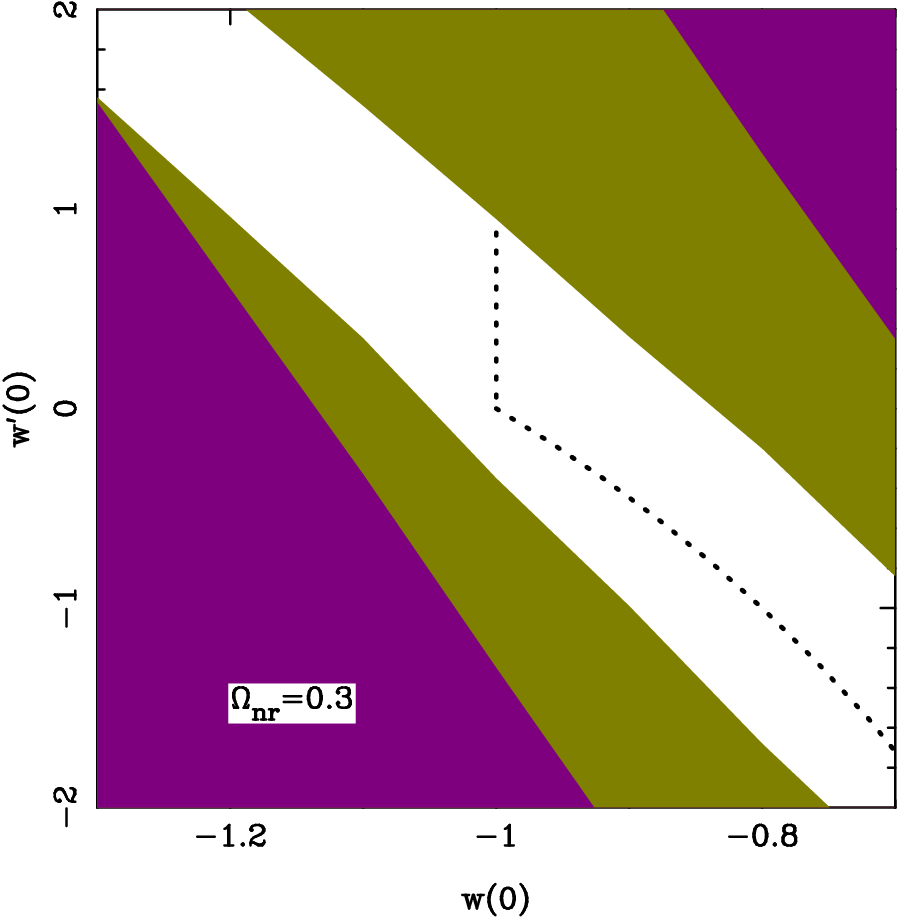}
\end{center}
\caption{The top panel shows the allowed region at $99\%$ confidence level in
  the $w'(0)-w(0)$ plane for $\Omega_{nr} = 0.3$ for the parameterization in
  Eq.(\ref{taylor}) with $p=1$.  
  The purple region is excluded by supernova observations. 
  WMAP constraints rule out the region in green at the same confidence level.
  The thick dotted line divides models that violate the strong energy
  condition from those that do not. 
  Models which preserve strong energy condition are on the top right of the
  dotted lines.  
  The lower panel shows the same plot for $p=2$.} 
\label{fig:w1w0}
\end{figure}

\section{Results}

Analysis of the high redshift supernova data shows that a large region in the
parameter space $\Omega_{nr} - w_0 - w_1$ is allowed at the $99\%$ confidence
level. 
For $w_1=0$, the favoured values of $w_0$ decreases (is more negative) with
increasing $\Omega_{nr}$.  
We will show detailed plots only for $\Omega_{nr}=0.3$ and then discuss the
variation of allowed range of parameters with $\Omega_{nr}$.  
We have studied large regions in parameter space explicitly and have not
relied on any expansion around the best fit model.

A fairly rapid evolution in the equation of state is allowed, so much
so that $w(z) \geq -1/2$ at $z \geq 0.5$ is consistent with the
supernova observations.   
Fig. \ref{fig:w1w0} shows the allowed region at $99\%$ confidence level in the
$w_1 - w_0$ plane for $\Omega_{nr} = 0.3$, regions in the parameter space that
are ruled out are blanked out in purple. 
The $99\%$ confidence limit corresponds to a $\Delta\chi^2=11.3$
for the three parameters we have here. 
These results compare well with earlier studies that have addressed
the issue of dark energy.
We are not showing the full region allowed by supernova constraints as much of
it is ruled out by WMAP observations. 

{\it Adding WMAP constraints reduces the allowed region in parameter space in a
significant manner.}  
The green region is ruled out by WMAP observations at $99\%$ confidence level
in Fig.~\ref{fig:w1w0}.  
We have $\Delta \chi^2=13.28$ for both $p=1$ and $p=2$.
The orientation of the confidence level contours is consistent with the
assertion that CMB observations essentially constrain $w_{eff}$.  
Dotted lines within this region mark the dividing line between models for
which $w(z) \geq -1$ at all times (to the right and above the dotted line) and
models which have $w(z) < -1$ at some redshift. 
These lines separate models that violate the  condition $(\rho+p)\geq 0$ from
those which do not. 
Equations for these lines are:
\begin{eqnarray}
p=1: \nonumber \\
&& w(0) = -1 \;\;\;\;\;\;\;\;\;\;\;\;\;\;\;\;\;\;\;\; (w'(0) \geq 0) \nonumber \\
&& w(0) + w'(0) = -1 \;\;\;\;\;\;\; (w'(0) < 0) \nonumber \\
p=2:  \nonumber \\
&& w(0) = -1 \;\;\;\;\;\;\;\;\;\;\;\;\;\;\;\;\;\;\;\;\;\;\; (w'(0) \geq 0) \nonumber \\
&& w(0) + w'(0)/2 = -1 \;\;\;\;\;\;\; (w'(0) < 0) \nonumber 
\end{eqnarray}
Observations do not discriminate between these two types of models but we
have indicated this line for the benefit of those who have a strong
theoretical preference. 

In Fig.~\ref{fig:w1w0}, the top panel is for $p=1$ and the lower panel is for
$p=2$. 
It is clear that the key features are the same in two panels.
For both the parameterizations the region allowed by WMAP observations is much
smaller than that allowed by the supernova observations (and the region
occupied by non-phantom models is still smaller.) 
The lower bound from supernovae becomes stronger than the CMB constraint at
small $w_0$. 
The quantitative differences between the figures can be attributed to
difference in $w(z)$ at at a fixed $z$ in the two parameterizations.

\begin{figure}
\begin{center}
\includegraphics[width=8.0cm]{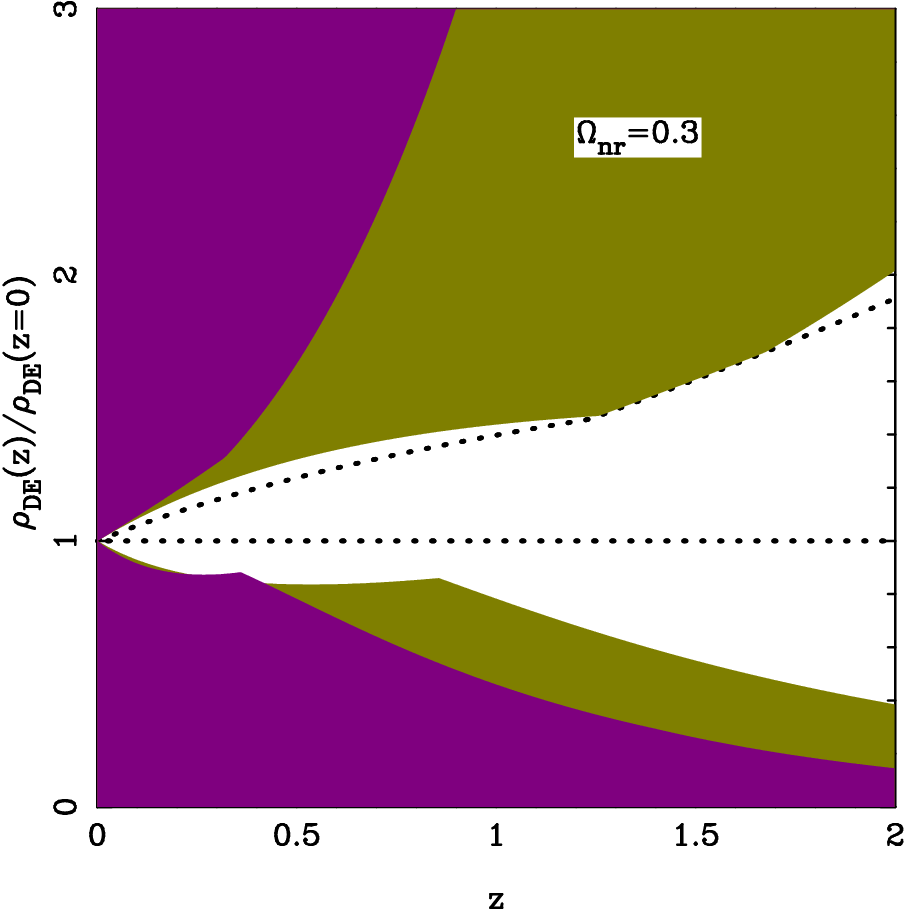}
\includegraphics[width=8.0cm]{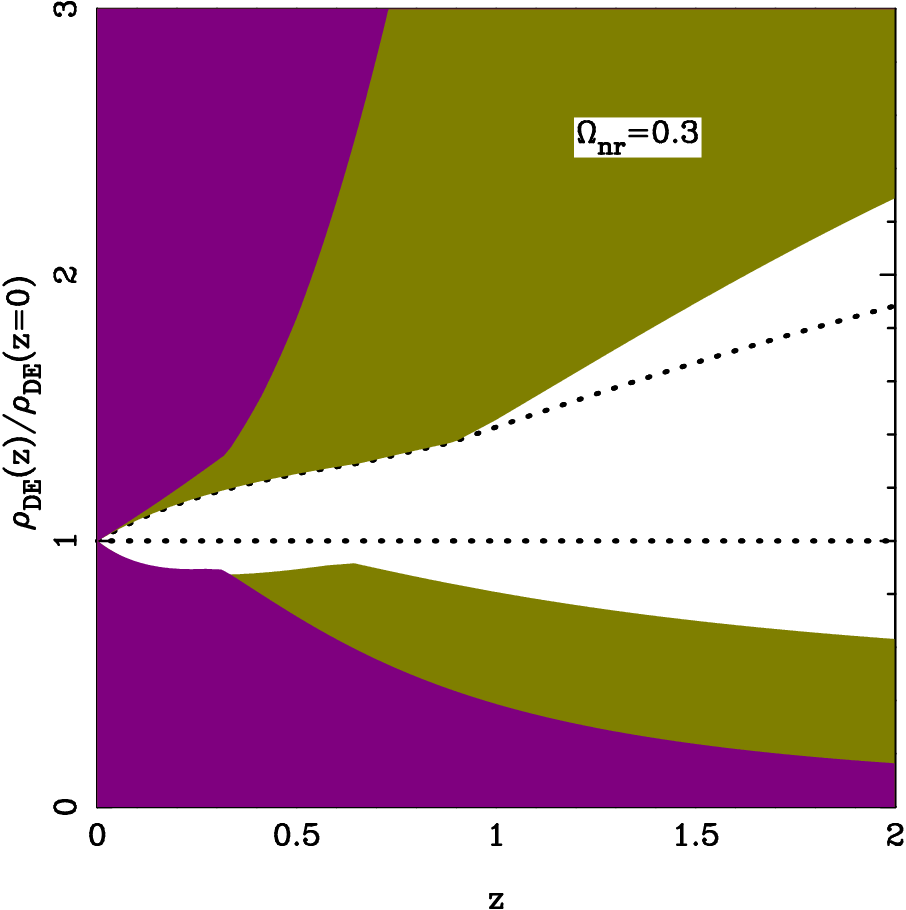}
\end{center}
\caption{The top panel shows those values of $\rho(z)$ for $2 \geq z
  \geq 0$ that are disallowed by the supernova and WMAP constraints
  for $\Omega_{nr}=0.3$.   
The purple region is ruled out by supernova observations and WMAP
constraints limit the allowed region by also ruling out the green
region.   
The remaining region contains allowed models that are parameterized by
Eq.(\ref{taylor}) with $p=1$.
The allowed region for models that do not violate the condition
$(\rho+p)\geq0$ is smaller and lies between the dotted lines.   
The lower panel shows the same plot for $p=2$.}
\label{fig:wenvelope}
\end{figure}

To get a better feel, we have shown the allowed values of $\rho^{DE}(z)$ for
dark energy as a function of redshift $z$ for $\Omega_{nr} = 0.3$ with just
supernova data and by combining the constraints from WMAP in
Fig.\ref{fig:wenvelope}.  
The purple region is ruled out by constraints from observations of high
redshift supernovae alone. 
In cases where the allowed region extended beyond the range of parameters
that we have studied, we used the largest allowed value within the range we
had studied to get the envelope shown for this plot.   
For example,  the allowed values of $w'(0)$ for supernova constraints go beyond
the range that we have studied ($-2 \leq w'(0) \leq 2$).
This leads us to underestimate the region allowed by supernova constraints
in fig.\ref{fig:wenvelope} {\it but this does not affect the WMAP
  constraints}.   
In this sense, our main conclusion --- that WMAP observations put stronger
constraint on $w(z)$ ---  is better than it appears from this figure.
The allowed variation in $\rho^{DE}(z)$ is slower than that
for matter, but considerable evolution is allowed by the supernova
observations. 
The green region is excluded by WMAP constraints.
For clarity, we have only marked those regions that are not already excluded
by the supernova observations. 
Models that never have $w(z) < -1$ live between the broken lines.
Models that are above the upper broken line aer typically those that have
$w(0) < -1$ but a positive $w'(0)$. 
Of course, a model that lives in the allowed region but has a different
functional form  for $w(z)$ from that in  Eq. (\ref{taylor}) may not be
allowed. 
The results for two parameterizations are very similar for $z \leq 2$.  

\begin{figure}
\begin{center}
\includegraphics[width=8.0cm]{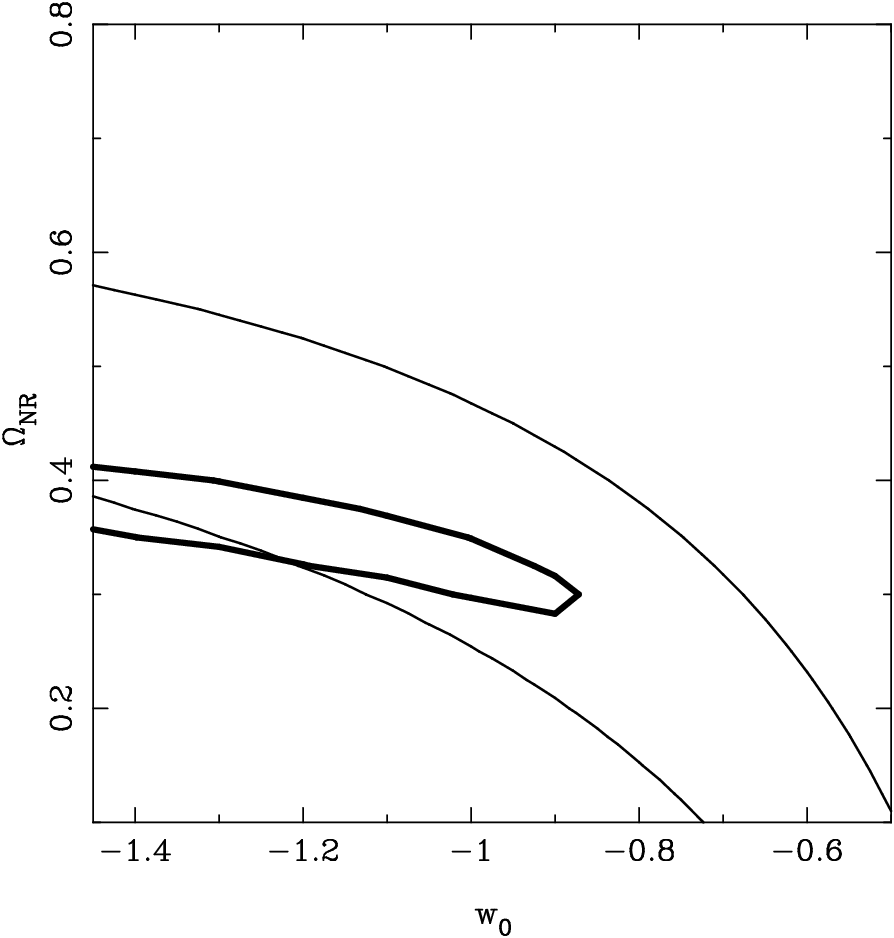}
\end{center}
\caption{The figure shows confidence levels in $\Omega_{nr}-w_0$
  plane for supernova and WMAP constraints (for both parameterisations) with
  $w'(z=0)=0$.  
 $99\%$ confidence levels are plotted for both observational constraints. The
  bold solid contour 
  is  $99\%$ confidence level for WMAP.  It is clear that the
  region allowed by the combination of two observational constraints is much
  smaller than the one allowed by supernova data. }
\label{fig:w0omega}
\end{figure}

\section{Conclusions}

The observations we have considered constrain distances and the main
constraint from CMB observations is on a reduced quantity $w_{eff}$ which is
the integrated effective value for the angular diameter distance. 
If $w(0) < w_{eff}$ ($w(0) > w_{eff}$) then $w'(0) > 0$ ($w'(0) < 0$) is
expected, indeed this explains the orientation of the allowed region in
figure~\ref{fig:w1w0}.
There is a strong degeneracy between $w_{eff}$ and $\Omega_{nr}$, $\partial
w_{eff}/\partial \Omega_{nr} < 0 $ (see Figure \ref{fig:w0omega}).
The cosmological constant itself is ruled out at $99\%$ confidence level for
$\Omega_{nr} > 0.375$ and only phantom models survive beyond this, thus if
other observational constraints were to rule out $\Omega_{nr} \leq 0.375$ then
we will be forced to work with phantom models.   
(Note that the Cosmological constant and other non-phantom models are not
ruled out by supernova observations for these values of $\Omega_{nr}$.)  
But with current observations, such a constraint does not exist and the
cosmological constant model is allowed.
Hence observations do not {\it require} varying dark energy or phantom models
even though such models are consistent with observations. 
The detailed dependence on $\Omega_{nr}$ and other parameters will be explored
in detail in a later work \citep{jbp2}. 

In our analysis we have not taken into account any perturbations in dark
energy.   
This effect becomes increasingly important as $w$ approaches zero, leading to
suppression of perturbations in matter
\citep{quint_growth1,quint_growth2,quint_growth3}.    
This effect is relevant only at $l \leq 20$ in most models and hence its
relative importance is small as we are using the full WMAP angular power
spectrum here.  
Perturbations in dark energy also have a non-adiabatic component, but this
cannot be incorporated in our analysis as it requires a detailed model for
dark energy whereas we are working with parameterised variations. 
We are carrying out a more detailed analysis where we allow parameters
like $n$, $h$, $\Omega_B$, $\tau$, etc. to vary. 
Preliminary results suggest that after marginalizing over other parameters the
region in the $w(0)-w'(0)$ allowed by CMB observations increases, mainly on
the lower side of the region shown in figure~\ref{fig:w1w0}.  
Supernova observations then provide stronger constraints for smaller $w'(0)$
while the CMB observations continue to provide much stronger constraints on
large $w'(0)$.  
Adding other observational constraints from structure formation reduces the
allowed region significantly as extreme variations of parameters allowed by
WMAP data predict too much or too little structure formation \citep{jbp2}. 
These observations are not very relevant in the present work as we are keeping
most parameters fixed.

Our analysis of models and observations shows that the allowed variation of
$\rho^{DE}(z)$ is strongly constrained by a combination of WMAP
data and observations of high redshift supernovae.
Indeed, given the allowed window as seen in Fig.~\ref{fig:wenvelope}, the
cosmological constant seems to be the most attractive dark energy candidate. 

\section*{Acknowledgements}

All the numerical work in this paper was done using cluster computing
facilities at the Harish-Chandra Research Institute
(http://cluster.mri.ernet.in/).

\label{lastpage}

\end{document}